%% file: main.tex
\documentclass[runningheads]{llncs}
\usepackage[utf8]{inputenc}
\usepackage[T1]{fontenc}
\usepackage{tabularx,longtable,afterpage}
\usepackage{url,caption,subcaption}
\usepackage{graphicx,adjustbox}
\usepackage{comment}
\usepackage{array,multirow}
\usepackage{amsmath,amssymb}
\usepackage{pdfpages,lscape}
\usepackage[english]{babel}

\begin{document}

\title{Effective Anonymous Messaging: \\ the Role of Altruism}
\titlerunning{Effective anonymous messaging}
\author{Marcell Frank\inst{1} \and Bal{\'a}zs Pej{\'o}\inst{1,2} \and Gergely Bicz{\'o}k\inst{1,2,3}}
\authorrunning{Frank, Pej{\'o}, Bicz{\'o}k}

\institute{CrySyS Lab, Dept. of Networked Systems and Services,\\ Budapest Univ. of Technology and Economics, Hungary\\
\email{marcell.frank@edu.bme.hu, \{pejo,biczok\}@crysys.hu}
\and
BME-HUN-REN Information Systems Research Group, Hungary
\and
University of Michigan, Ann Arbor, MI 48109, USA
}

\maketitle          

\begin{abstract}
    Anonymous messaging and payments have gained momentum recently due to their impact on individuals, society, and the digital landscape. Fuzzy Message Detection (FMD) is a privacy-preserving protocol where an untrusted server performs message filtering for its clients in an anonymous way. To prevent the server from linking the sender and the receiver, the latter can set how much cover traffic they should download along with genuine messages. Clearly, this could cause unwanted messages to appear on the user’s end, thereby creating a need to balance one’s bandwidth cost with the desired level of unlinkability. 

    Previous work showed that FMD is not viable with selfish users.
    In this paper, we model and analyze FMD using the tools of empirical game theory and show that the system needs at least a few altruistic users to operate properly.
     Utilizing real-world communication datasets, we characterize the emerging equilibria, quantify the impact of different types and levels of altruism, and assess the efficiency of potential outcomes versus socially optimal allocations. Moreover, taking a mechanism design approach, we show how the betweenness centrality (BC) measure can be utilized to achieve the social optimum.
\end{abstract}

\keywords{Anonymous Messaging \and Privacy \and Fuzzy Message Detection \and Altruism \and Game Theory \and Best-Response Dynamics}

\input{1intro}

\input{2background}
\input{3model}
\input{4implementation}

\input{5result}
\input{6conclusion}

\section*{Acknowledgements}
Project no. 138903 has been implemented with the support provided by the Ministry of Innovation and Technology from the NRDI Fund, financed under the FK\_21 funding scheme.

\bibliographystyle{splncs04}
\bibliography{reference}


\end{document}

%% file: 1intro.tex
\section{Introduction}
\label{sec:intro}

Anonymous messaging is a critical enabler in the landscape of digital privacy, as it allows individuals to send and receive information without revealing their identities. By doing so, it ensures a degree of confidentiality by providing freedom of expression and freedom of association. Such mechanisms foster trust and autonomy for users seeking advanced confidentiality in their communications and financial transactions. Anonymous messaging is realized by various cryptographic protocols, which offer a shield against surveillance and unauthorized access. One acclaimed cryptographic solution is Fuzzy Message Detection (FMD)~\cite{beck2021fuzzy}. 

In a fully relationship-anonymous setup, even intended recipients remain unaware of messages sent to them without decrypting the entire traffic, causing computational inefficiency and wasting bandwidth. Indeed, if messages (transactions) are continuously posted to a public board (e.g., a permissionless blockchain ledger), the user (with limited resources) must scan the entire chain to pick the messages intended for them.

\begin{figure}[t]
    \centering
    \includegraphics[width=.75\textwidth]{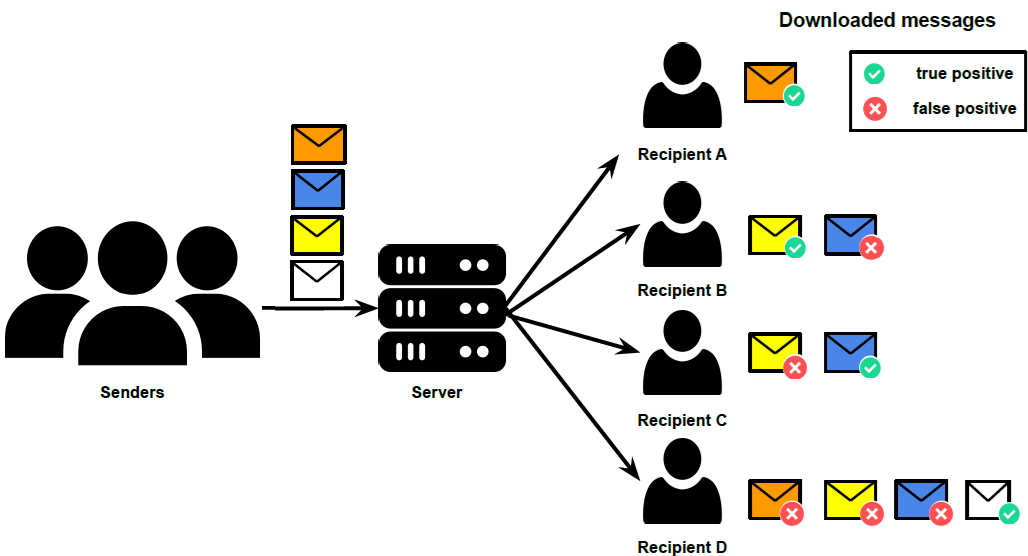}
    \caption{FMD failure: Recipient D has seemingly maximum protection because it downloads all messages as cover traffic, yet its genuine message (white envelope) is not downloaded by any other participants, so no relationship anonymity is provided.}
    \label{fig:FMD}
\end{figure}

FMD is a relatively new privacy-enhancing cryptographic technique with several desired privacy properties, such as relationship anonymity (i.e.,  unlinkability). FMD provides a workaround by enabling users to delegate the detection of incoming traffic to an untrusted server in an efficient and privacy-hardened way. It allows users, when online, to download a mixed set of messages in which some are addressed to the user some to others, based on their chosen false-positive detection rate. The cryptographic method ensures that the server cannot distinguish between true and false-positive messages, effectively using the latter as cover traffic. The FMD protocol is illustrated in Fig.~\ref{fig:FMD}. 

This promising technique has garnered attention for its adaptability in various scenarios; see, e.g., the Niwl anonymous messaging app~\cite{lewis2021niwl}, which planned to implement FMD. Concerning anonymous payments, there have been efforts to incorporate FMD into privacy-preserving cryptocurrencies (such as Penumbra~\cite{devalence2021penumbra}) and into privacy-enhancing overlays (such as Zeth~\cite{rondelet2021zeth}). However, since the initial hype, none of these use cases appear to have come through. (We hope our results can restart the discussions around the real-life viability of FMD.)

Additionally, despite its seemingly attractive properties and fleeting commercial interest, the privacy protection that FMD provides is far from air-tight. Seres et al. showed that statistical attacks can break FMD's guarantees concerning relationship anonymity, recipient unlinkability, and temporal detection ambiguity~\cite{seres2022effect}. In terms of relationship anonymity, they also showed that selfish users had no incentive to maintain non-zero cover traffic, as it is cost-bearing and their own protection level is independent of it; see Fig.~\ref{fig:FMD}.

This scenario falls under the tragedy of the commons~\cite{hardin_toc}, in the sense that every user benefits from consuming a public good (i.e., privacy) but is not willing to contribute to it (also referred to as free-riding). Many socio-technical systems were shown to exhibit this type of behavior, from peer-to-peer file-sharing~\cite{freeriding_p2p} through collaborative physical and cybersecurity~\cite{kunreuther2003interdependent,laszka_interdepsec} to pandemic response measures~\cite{pejo_covid}. The economic literature proposes the internalization of externalities and/or appropriate regulatory measures to resolve such situations; however, altruistic behavior also has the potential to alleviate the ineffective equilibrium~\cite{altruism1,altruism2}.

We believe that altruistic behavior is especially realistic in privacy-preserving communications, where the actual stakeholders are often members of the same community; it has been shown that individuals in tightly-knit groups (and societies with a strong sense of duty) routinely choose to act for the good of others at a cost to themselves~\cite{DBLP:conf/ijcai/Yu0V21}. Interestingly, awareness-raising campaigns (e.g., related to privacy around the introduction of the GDPR or social distancing during the COVID-19 pandemic) try to make people internalize their externalities, changing their mental models~\cite{mental_idp}.

\noindent\textbf{Our contribution. }
In this paper, we investigate and quantify the impact of altruism in anonymous messaging networks. Specifically, we seek answers to the following research questions:

\begin{itemize}
    \item[RQ1] How do the type of altruism, the number of altruistic players, and the network topology affect the equilibrium outcome?
    
    \item[RQ2] How can a central planner (e.g., messaging app provider) set the false positive rates to achieve social optimum? Is there an easily computable metric that can be an efficient proxy for the optimal false positive rate in a realistic setting?
    
    

\end{itemize}

Our results advance the state of the art in multiple aspects, as we made the following contributions. 

\begin{itemize}
    \item Adhering to the principles of empirical game-theoretic analysis~\cite{DBLP:conf/aaai/Wellman06}, we a) focus on the FMD technology to construct meaningful utility functions, b) use real-world datasets of communication networks and patterns, and c) employ heuristic methods to analyze game outcomes.
    \item We extend the selfish game in~\cite{seres2022effect} with different notions of altruism, where some nodes care about the welfare of their neighbors (local altruism) or all other nodes (global altruism). 
    \item We show that a) the system reaches a viable, non-trivial equilibrium with only a few altruistic nodes, and b) we characterize emerging equilibria with respect to efficiency and the impact of different types and levels of altruism. 
    \item We find that a central mechanism designer (e.g., the developer/maintainer of a messaging app) could use betweenness centrality as a proxy metric for assigning optimal false positive rates.    
\end{itemize}

The rest of the paper is organized as follows. Section~\ref{sec:back} summarizes the relevant background and related work. Section~\ref{sec:modeling} defines the altruistic game model. Section~\ref{sec:imp} describes our analysis approach. Section~\ref{sec:results} presents our findings. Finally, Section~\ref{sec:conc} outlines future work and concludes the paper. Note that an extended version of our paper is available online\footnote{\url{https://cloud.crysys.hu/s/fmdgt}}~\cite{tech}.

%% file: 2background.tex
\section{Background}
\label{sec:back}

Here, we establish the preliminaries regarding game theory and the FMD mechanism and give a brief overview of related work.

\subsection{Preliminaries}

\vspace{0.1cm}\noindent\textbf{Game Theory. }
A non-cooperative game model consists of players, strategies, utility functions, and the mechanics of game playing. In this paper, we study one-shot games, where the complexity comes from the large number of players and the underlying network structure.

A Nash Equilibrium (NE)~\cite{nash_gt_original} occurs when none of the players can unilaterally deviate from their chosen strategy without incurring lower utility. Notably, such NE always exists if each player can choose from a finite set of actions. In contrast, the Social Optimum (SO) is the set of strategies where the overall utility of all the agents is maximized. The ratio between this and the worst and best NE is called Price of Anarchy (PoA)~\cite{article:priceofanarchy} and Price of Stability (PoS)~\cite{article:priceofstability}, respectively. These benchmarks express how the overall system performance degrades due to the selfish behavior of its agents.

Relaxations of both NE and SO exist, which are computationally more feasible to obtain. Within the paper, we utilize the $\varepsilon$-Equilibrium concept~\cite{roughgarden_2016}, where agents may gain a limited utility by deviating from their current strategy. Similarly, we define $\varepsilon$-SO, where no user's strategy could be changed in a way that would result in a larger than $\times(1-\epsilon)$ overall improvement.
One way how such equilibria might be found is via the Best Response Mechanism (BRD)~\cite{roughgarden_2016}, where the players are iteratively changing their actions to maximize their payoffs. In particular, if the game is a potential game~\cite{article:Shapley1994PotentialG}, i.e., the incentive of all players to change their strategy can be expressed using a single function, then the BRD is ensured to converge to an NE.

Finally, altruism refers to a player's willingness to incur personal costs to benefit others, even when it conflicts with their self-interest~\cite{simon1993altruism}. It involves acting for the greater good, potentially leading to cooperative behavior that can influence outcomes in strategic interactions~\cite{fletcher2007evolution}. 


\vspace{0.1cm}\noindent\textbf{Fuzzy Message Detection. }
In a sense, Fuzzy Message Detection (FMD)~\cite{beck2021fuzzy} is an extension of asymmetric encryption, where the public keys are replaced with so-called ``detection keys''. In the classical setup, the users share with the server their public keys, and the server sends back to them the ciphertexts, which are encrypted with them. In contrast, besides matching with genuine ciphertexts, detection keys would also match other ciphertexts (encrypted with different public keys). This way, the genuine and the cover traffic would be indistinguishable for the server without access to the private key. Consequently, the server may send the same ciphertext to several users, unaware of who was the originally intended recipient. Yet, besides the genuine messages, the clients cannot decrypt other messages, as they do not hold the appropriate private keys. Hence, their sole purpose is to provide cover traffic. As such, FMD prevents the leakage of metadata to some extent by creating ambiguity in the server concerning the destination of each message. 

The amount of cover traffic is determined by the false positive detection rate corresponding to each detection key. It determines the probability that a single non-matching ciphertext will be ``detected'' as matching. 
Formally, FMD provides Correctness (every message reaches its intended target), Fuzziness (targets receive additional messages proportional to their false positive detection rate), and Detection Ambiguity (only the targets can distinguish between genuine and cover messages). Through this paper, we follow the author's recommendation (regarding the efficiency of implementation) and set all false positive detection rates to be a power of two. We refer to Appendix B of our technical report~\cite{tech} for further details. 

\subsection{Related Work}

\noindent\textbf{FMD Alternatives. }
Since the introduction of FMD, a handful of works have attempted to tackle similar problems, such as Private Signaling (PS)~\cite{madathil2022private} or Oblivious Message Retrieval (OMR)~\cite{liu2022oblivious}. Other related problems were studied within the Private Information Retrieval (PIR)~\cite{chor1998private} literature. Note that this list is not exhaustive; we merely want to indicate our analysis may also generalize to other systems.

PS provides recipient privacy and key unlinkability, but its constructions rely upon strong environmental constraints, such as trusted hardware and two communicating but non-colluding servers. Although a recent work~\cite{jakkamsetti2023scalable} improved its scalability, trusted hardware is still assumed. OMR provides denial-of-service resistance besides the previously mentioned properties but comes with a heavy computational burden. Although a recent work~\cite{liu2023group} extended OMR to group messages, the computational burden only increased. 
Although our analysis is specific to FMD, the game-theoretic framework could be adapted to other anonymous messaging protocols or even generalized further; see Section~\ref{sec:conc} for details.

\vspace{0.1cm}\noindent\textbf{Free-riding in distributed systems. }
 The free-riding problem, emerging inefficient equilibria, and potential remedies have been studied extensively in distributed systems. One of the most scrutinized domains in this aspect is peer-to-peer systems
 ~\cite{feldman2006free}. In fact, the impact of disincentivized nodes was investigated in multiple real-world systems such as Gnutella~\cite{freeriding_gnutella}, Napster~\cite{freeriding_napster}, and BitTorrent~\cite{jun2005incentives}. Another much-researched domain, where the public good to be consumed is physical or cybersecurity, is interdependent security. Starting from the seminal works of Kunreuther and Heal~\cite{kunreuther2003interdependent} and  Varian~\cite{varian2004system}, there has been a line of research on interdependent security games~\cite{laszka_interdepsec}. Furthermore,  falling closer to our work, contributor incentives have been taken into account in the design of the Tor anonymous communication network~\cite{tor_incentives}. Specifically, the balance between bandwidth cost and privacy protection was studied in~\cite{gpath_tor}.

\vspace{0.1cm}\noindent\textbf{FMD Analysis. }
This paper was inspired by Seres et. al.~\cite{seres2022effect}, which lays the preliminary groundwork for studying anonymous messaging through the lens of game theory. The authors studied FMD from multiple angles and concluded it performs weakly in nearly all privacy aspects. Specifically, they assumed selfish participants and showed that setting the false positive detection rates to zero is an NE, which rendered the entire FMD protocol useless.
Their analysis did not consider altruism and assumed homogeneous users with random false positive detection rates.

 
In this paper, we study the impact of altruistic nodes in the FMD anonymous messaging system, where altruism invokes higher bandwidth costs corresponding to cover traffic. In fact, our results show a phenomenon similar to \cite{varian2004system}: the effort induced in equilibrium is highly concentrated at key nodes while others contribute little; yet, the system is functional as opposed to one with only selfish participants~\cite{seres2022effect}.



%% file: 3model.tex
\section{Model}
\label{sec:modeling}

In this section, we recap the game-theoretic model of the FMD anonymous messaging system introduced in~\cite{seres2022effect} and extend it with altruism. Following~\cite{seres2022effect}, we denote with $u$ the users and the number of their genuine incoming messages with $in_u$. The total number of messages in the system is $M$ while the false positive detection rate of $u$ is $p_u$, which implies that the expected number of messages assigned to $u$ by the server is $in_u+p_u \cdot (M-in_u)$. 

\subsection{The Selfish Game}
\label{sec:gamedef}

Seres et al.~\cite{seres2022effect} also defined $\alpha_u$ as the event of a relationship anonymity breach caused by a single message, where the server can link the known sender to recipient $u$. This event occurs if no other user downloads that particular message providing cover traffic, with a probability of $\alpha_u=\prod_{v\in\mathcal{N}/\{u\}} (1-p_v)$. Consequently, the probability of a breach from any incoming message is $1-(1-\alpha_u)^{in_u}$, i.e., the complement of a single breach happening but for all of the incoming messages.

Using this quantity, they defined the FMD game, where the players are the participants in the FMD protocol, and their strategies are their false positive detection rates (corresponding to the amount of their generated cover traffic). 
The game focuses on relationship anonymity, i.e., a privacy breach occurs when the server learns that two users are indeed communicating.

\begin{definition}[FMD Game]
    The FMD Game is a tuple $\langle\mathcal{N},\Sigma,\mathcal{U}\rangle$, where the set of players is $\mathcal{N}=\{1,\dots,U\}$, their actions are $\Sigma=\{p_1,\dots,p_U\}$ where $p_u\in\{2^{-1},2^{-2},\dots,2^{-10},0\}$ for $1\le u\le U$, and their utility functions are $\mathcal{U}=\{\varphi_u(p_1,\dots,p_U)\}_{u=1}^U$ such that for $1\le u\le U$:
    \begin{equation}
        \label{eq:game_basic}
        \varphi_u(\cdot)=-\underbrace{L\cdot(1-(1-\alpha _u)^{in_u})}_{C_u^{P}}-\underbrace{f\cdot(in_u+p_u\cdot(M-in_u))}_{C_u^{BW}}
    \end{equation}
    where $L$ is the cost of a privacy breach, and $f$ is the bandwidth cost of retrieving a single message from the server.
\end{definition}

To ease readability, we denote the first privacy-related expression of the equation as $C_u^{P}$ and the second bandwidth-related part as $C_u^{BW}$. One would expect a clear trade-off between privacy and bandwidth efficiency; however, Equation~\ref{eq:game_basic} highlights that a larger false-positive rate $p_u$ corresponds only to higher bandwidth (as more messages need to be downloaded from the server), but not (necessarily) to lower privacy loss. Indeed, upon closer inspection, it can be seen that $C_u^{P}$ is independent of the user's own action $p_u$. In fact, this renders the entire FMD protocol obsolete, as no rational user would opt-in to utilize any cover traffic. 

\begin{theorem}[Seres, Pej{\'o}, and Burcsi~\cite{seres2022effect}]
    The only NE for the FMD Game is where no one utilizes any cover traffic, i.e., $p_u=0$ for all $1\le u\le U$.
\end{theorem}


\subsection{The Altruistic Game}
\label{sec:altruism}

As a consequence of this simple theorem, FMD is not viable with only selfish users and no incentive re-design. (Note that the latter could take the form of payments or rewards, similar to peer-to-peer systems~\cite{p2p_incentives_survey2023} or recent federated learning schemes~\cite{huang2020exploratory}. This direction could be important for future work.)

However, the presence of altruistic users could change the game (both literally and metaphorically).
As pointed out in Section~\ref{sec:intro}, it is plausible among privacy-conscious individuals in the same community, i.e., users of the same messaging app, to behave altruistically~\cite{DBLP:conf/ijcai/Yu0V21}. We consider altruism in the form of extending the utility function to encapsulate the selflessness of the players; they could act in a way that benefits others at a cost to themselves. 

\begin{definition}[Altruistic player]
    A player is altruistic if its utility is directly affected by the welfare of others.
\end{definition}

As opposed to~\cite{article:selfishness-level}, where the entire social welfare is appended to the utility function of the altruistic players, we add only the privacy loss $C_u^P$ of other players as the motivating factor behind altruism (and the community effect) is privacy itself. This third term in the utility function is added through a multiplicative factor referred to as \emph{altruistic constant $a_u$}, indicating the level of altruism (or selfishness) of the respective user. For a selfish player $u$, $a_u=0$. On the other hand, if $u$ is altruistic, this value would be positive, i.e., $a_u>0$. In a sense, $a_u$ captures a player's willingness to cooperate for the greater good (social welfare). An intuitive expectation is that ''enough`` altruism would shift the selfish NE towards the SO~\cite{article:selfishness-level}.

The FMD game is played on a communication graph where a weighted directed edge $e(u,v)$ exists if user $u$ sends at least one message to user $v$. We consider the network topology and the corresponding communication pattern as given: no strategic decisions are made regarding network formation.

In the context of anonymous messaging (and potentially other applications involving an underlying network), altruism itself could have multiple meanings. We consider two kinds of altruism: local (i.e., caring about the welfare of your contacts) and global (i.e., caring about the welfare of the whole society). Although there could be other alternative interpretations, we believe these two have intuitive and inherent significance in our application scenario.


\vspace{0.1cm}\noindent\textbf{Local Altruism. }
In the context of FMD, local altruism pertains to cost-bearing actions that improve the welfare of directly connected nodes. 

\begin{definition}[L-FMD Game]
    \label{def:L}
    The L-FMD Game extends the FMD game with the cost of local altruism in the utility function for $1\le u\le U$:
    \begin{equation}
        \label{eq:local}
        \varphi_u(\cdot)=-C_u^{P}-C_u^{BW}-\underbrace{a_u\cdot \sum_{v:v\sim u}C_v^{P}}_{C_u^{LA}}
    \end{equation}
    where $a_u$ is the altruistic constant and $v\sim u$ means that user $v$ has a connection with user $u$, i.e., there is message flow (in any direction) between them. The direction is relaxed as the model regards relationship anonymity (transitive).   
\end{definition}

\vspace{0.1cm}\noindent\textbf{Global Altruism. }
In the context of FMD, global altruism acknowledges other-regarding behavior affecting the welfare of any node in the communication network. Such behavior recognizes the ultimate interdependence of online privacy~\cite{biczok2013interdependent}.

\begin{definition}[G-FMD Game]
    \label{def:G}
    The G-FMD Game extends the FMD game with the cost of global altruism in the utility function for $1\le u\le U$:
    \begin{equation}
        \label{eq:global}
        \varphi_u(\cdot)=-C_u^{priv}-C_u^{BW}-\underbrace{a_u\cdot \sum_{v\in\mathcal{N}/\{u\}}C_v^{P}}_{C_u^{GA}}
    \end{equation}
\end{definition}

Intuitively, altruistic players may be able to compensate for the lack of cover traffic from selfish nodes by setting their false positive detection rate high, thereby improving the privacy of their immediate neighborhood (local) or the whole society (global).



%% file: 4implementation.tex
\section{Experiments}
\label{sec:imp}

As altruistic FMD games do not lend themselves easily to theoretical analysis and we wanted to quantify the effect of various system parameters, we took an empirical approach~\cite{DBLP:conf/aaai/Wellman06}. Here, we detail i) the real-world communication pattern datasets used, ii) the best-response dynamics (BRD) algorithm implemented, and iii) the importance of choosing the initial candidate strategy distribution for the BRD. Our code, datasets, and results are available online\footnote{\url{https://github.com/m9framar/FMD-GT}}.

\subsection{Datasets}

We simulate FMD game instances on data from real messaging systems\footnote{\url{http://snap.stanford.edu/temporal-motifs/data.html}}. 
The first is the College Instant Messaging dataset~\cite{collegemsg}, referred to as \textit{message}; it contains the instant messaging network of college students from the University of California, Irvine. The graph consists of $1,899$ nodes (students) and $59,835$ edges (messages) spanning $193$ days. The second one is the EU E-mail dataset~\cite{eumessage}, denoted as \textit{mail}; it contains a collection of emails between members of a large European research institution. The network consists of $986$ nodes (researchers) and $332,334$ edges (emails) over $803$ days. Note that \textit{mail} represents a much denser communication network compared to \textit{message}.

While running the BRD algorithm, we realized that we had to ease the computational burden of our experiments. Therefore, we ``halved'' both graphs:  we ordered the nodes by degree and discarded every second node along with any edge connected. 
The corresponding statistics can be found in Appendix C of our technical report~\cite{tech}. We conjecture that betweenness centrality captures the ``importance'' of a node well in this context; we define this measure here for further use.

\begin{definition}[Betweenness Centrality (BC)~\cite{freeman1977set}]
    The betweenness centrality for each vertex is the number of shortest paths (from the set of all possible shortest paths between all node pairs) that pass through the vertex. 
\end{definition}

\subsection{Best-response dynamics}

Applying the BRD to a potential game will always yield an NE; this was also true for the selfish FMD games, revealing a unique all-zero equilibrium~\cite{seres2022effect}. In contrast, the inclusion of altruism modifies the objective function such that it no longer constitutes a potential function. Moreover, the altruism term might add local minima to the objective function; thus,  the BRD might stop at multiple different equilibria, depending on the initial state.
In this paper, we utilize the $\varepsilon$-BRD~\cite{roughgarden_2016}, which works in a sequential manner where, at each step, a single player changes its strategy, and each time, its chosen strategy value (false-positive detection rate $p_u$) can be incremented (decremented).  To simplify computations, we discretize and bound the value set such that $p_u \in \{0,2^{-10},\dots,2^{-1}\}$, in accordance with the recommendation of the creators of FMD~\cite{beck2021fuzzy}. For simplicity, we also restricted the possible values for the altruistic constant $a_u$; in a single experiment, all altruistic actors are characterized by either $a_u=0.1$ or $a_u=1$, respectively. We set $\varepsilon = 10^{-5}$ for all experiments. Note that both L-FMD and G-FMD are defined as one-shot games; the BRD algorithm is just a tool to find the $\varepsilon$-NE. 
 
\begin{definition}[$\varepsilon$-BRD (Maximum Gain)~\cite{roughgarden_2016}]
    This algorithm is a slight modification of the original BRD, where in one iteration, only a single node (the one corresponding to the highest utility gain) updates its strategy with a single increment/decrement until no player can increase its payoff by at least $\varepsilon$. 
\end{definition}

\subsection{Initial strategy candidates for BRD}

We experimented with various initial false positive detection rate settings to ensure the comprehensive exploration of the search space and find all possible $\varepsilon$-NE for both altruism types. We also re-used the same approach to establish the social optimum. We used three different strategies for initial settings.


\begin{enumerate}
    \item \textbf{Thresholding}: players' initial candidate strategy depends on a predetermined value of a node property, either betweenness centrality or degree number. The false positive detection rate is set to either $2^{-1}$ or $2^{-10}$ for the nodes above the threshold. Note that a zero threshold still allows for $0$ initial false positive detection rates for nodes with 0 property values.       
    
    \item \textbf{Sorting}: players are assigned an initial candidate strategy based on their relative position in an ordered list according to a node property, either betweenness centrality or degree number. Nodes with the highest values are assigned $2^{-1}$, nodes with the lowest values are assigned $2^{-10}$, while the initial candidate strategies of in-between nodes are calculated based either on linear (equal cardinality of buckets) or exponential ``intrapolation'' (size of the buckets follow $[1,2,4,\dots]$, where the last bucket consists of the rest of the users).
    
    \item \textbf{Random}: similarly to~\cite{seres2022effect}, nodes are randomly assigned a possible strategy.
\end{enumerate}

We set the threshold for the normalized betweenness centrality to $0.01$ and for degree values to $4$. With these values the computational burden for the experiment was still manageable, while they landed themselves on non-trivial results. 

The idea behind the uniform initial strategies $2^{-1}$ and $2^{-10}$ for \emph{Thresholding}  is that $\varepsilon$-BRD could possibly reach the same final strategy distribution, but from the opposite extremum of the search space (a unique equilibrium for the given parameter settings). On the other hand, if they converge to different equilibria, that could form the basis for PoA and PoS calculations. The intuition behind \emph{Sorting} is the ``importance'' of nodes could be a good proxy for an efficient strategy profile.

%% file: 5result.tex
\section{Results}
\label{sec:results}

In this section, we first establish the social optimum of the selfish FMD game (not given in ~\cite{seres2022effect}), and then we study altruistic L-FMD/G-FMD games with respect to both equilibrium and social optimum.
For all experiments, we set the bandwidth cost of a single message to be $f=1$ and the privacy loss $L=|E|-\max_u[in_u]+1$, which yields $L_{\mathrm{college}}=14797$ and $L_{\mathrm{mail}}=77947$, respectively. As stated before, we used $\varepsilon=10^{-5}$ for the $\varepsilon$-BRD and studied the impact of altruism by varying the altruism type (local or global) and the altruism constant: $a_u \in \{0.0,0.1,1.0\}$. 

\subsection{Social Optimum}

\begin{figure}[!b]
    \centering
    \includegraphics[width=0.5\textwidth]{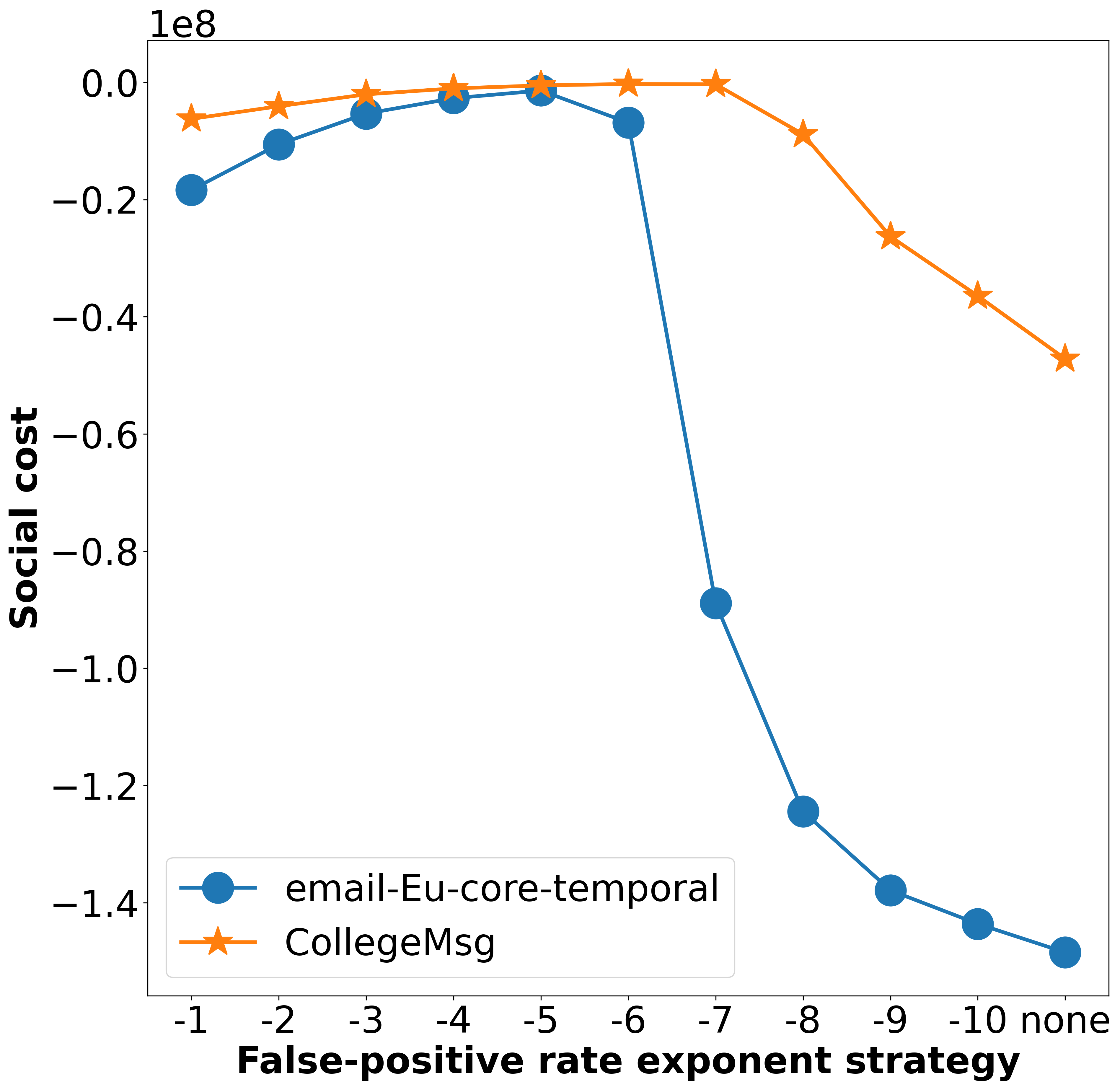}
    \caption{SO without altruism, uniform false positive rates.}
    \label{fig:swcentral}
\end{figure}

\vspace{0.1cm}\noindent\textbf{SO without altruism. }
To facilitate comparability with \cite{seres2022effect} and to provide an easily understandable baseline, we consider a simple scenario where a Mechanism Designer could only set the false positive rates in a uniform manner (same value for all nodes). Fig.~\ref{fig:swcentral} shows that the corresponding social welfare is the highest with $p=2^{-6}$ and $p=2^{-7}$ for the \textit{message} and \textit{mail} dataset respectively. It can be seen that larger cover traffic is optimal for the sparser graph; this is intuitive as it is easier to infer relationships with fewer nodes and communication flows. Note how applying zero cover traffic (which is the NE without altruism) corresponds to the lowest social welfare.

\vspace{0.1cm}\noindent\textbf{SO With Altruism. }
Concerning the altruistic game, the social welfare for the \textit{message} dataset is maximized with the strategy profile shown in Fig.~\ref{fig:college_SO}. The top row corresponds to global and the bottom to local altruism, with the left column $a_u=0.1$ and the right column $a_u=1.0$. It is visible that the more prominent the altruism, the more cover traffic is optimal in the system (from bottom to top and from left to right). Notably, the (local, $0.1$) setting yields an SO very close to the homogeneous selfish SO ($2^{-6}$ for all). Also, all nodes contribute in every setting.

Results are similar for the \textit{mail} dataset with peaks one step lower (not shown); this is in line with the homogeneous selfish SO ($2^{-7}$ for all). Moreover, all initialization strategies yielded near-identical optimal strategy profiles.

\subsection{Equilibrium Analysis}

Here, we present various $\varepsilon$-equilibria we found; while there are many potential equilibria in these large games, our extensive experiments enabled us to identify the \emph{types} of equilibria that emerge from altruistic FMD games. We discuss these through examples grouped by the BRD initialization strategy used to discover them, allowing us to reflect on the computational aspect organically. Curiously, the \emph{Sorting} initialization scheme resulted in NE that i) were not among the best or worst and ii) did not provide additional insights; hence, we omit them.

\vspace{0.1cm}\noindent\textbf{Random initialization. }
The BRD for the \textit{mail} dataset converges from a random distribution as shown in Fig.~\ref{fig:mail_NE}. The top row corresponds to global and the bottom to local altruism, with the left column $a_u=0.1$ and the right column $a_u=1.0$. Similarly to the SO, the resulting NE means more cover traffic when altruism is more prominent, i.e., global vs. local, and a higher altruistic constant $a_u=1.0$.
Local altruism encourages the majority of players not to contribute, while a small subset of nodes ($12$ or $16$, depending on the level of altruism) provides maximum cover traffic. Note that the global altruistic NE is closer to the corresponding SO (see Fig.~\ref{fig:college_SO}), still with some nodes providing maximum cover traffic but with all nodes contributing.

\afterpage{
\begin{figure}[!htb]
    \centering
    \includegraphics[width=0.85\textwidth]{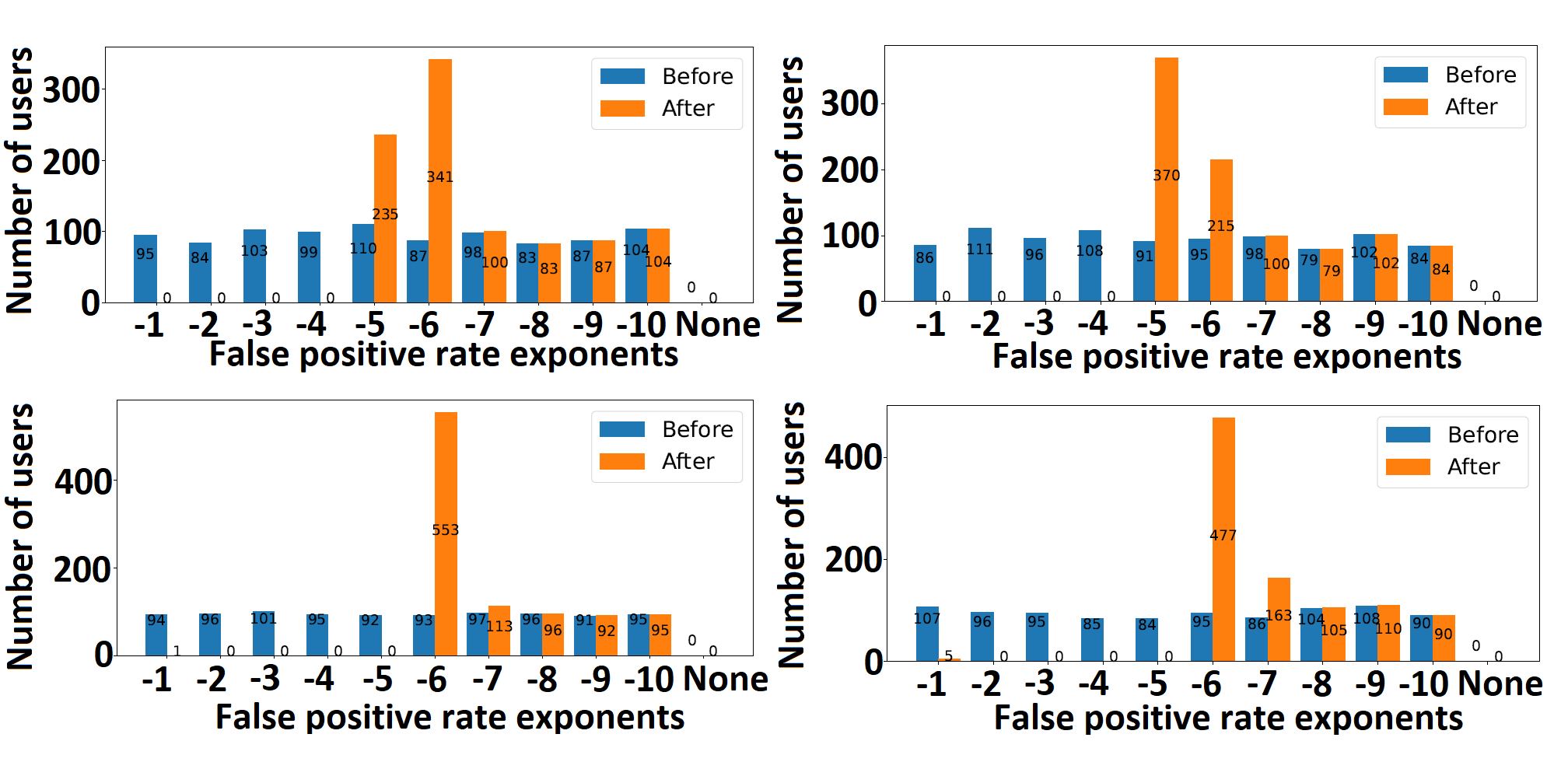}
    \caption{Strategy profile in SO, \textit{message} dataset, top: global altruism, bottom: local a., left: $a_u=0.1$, right $a_u=1.0$.}
    \label{fig:college_SO}
    \vspace{0.2cm}
    \includegraphics[width=0.85\textwidth]{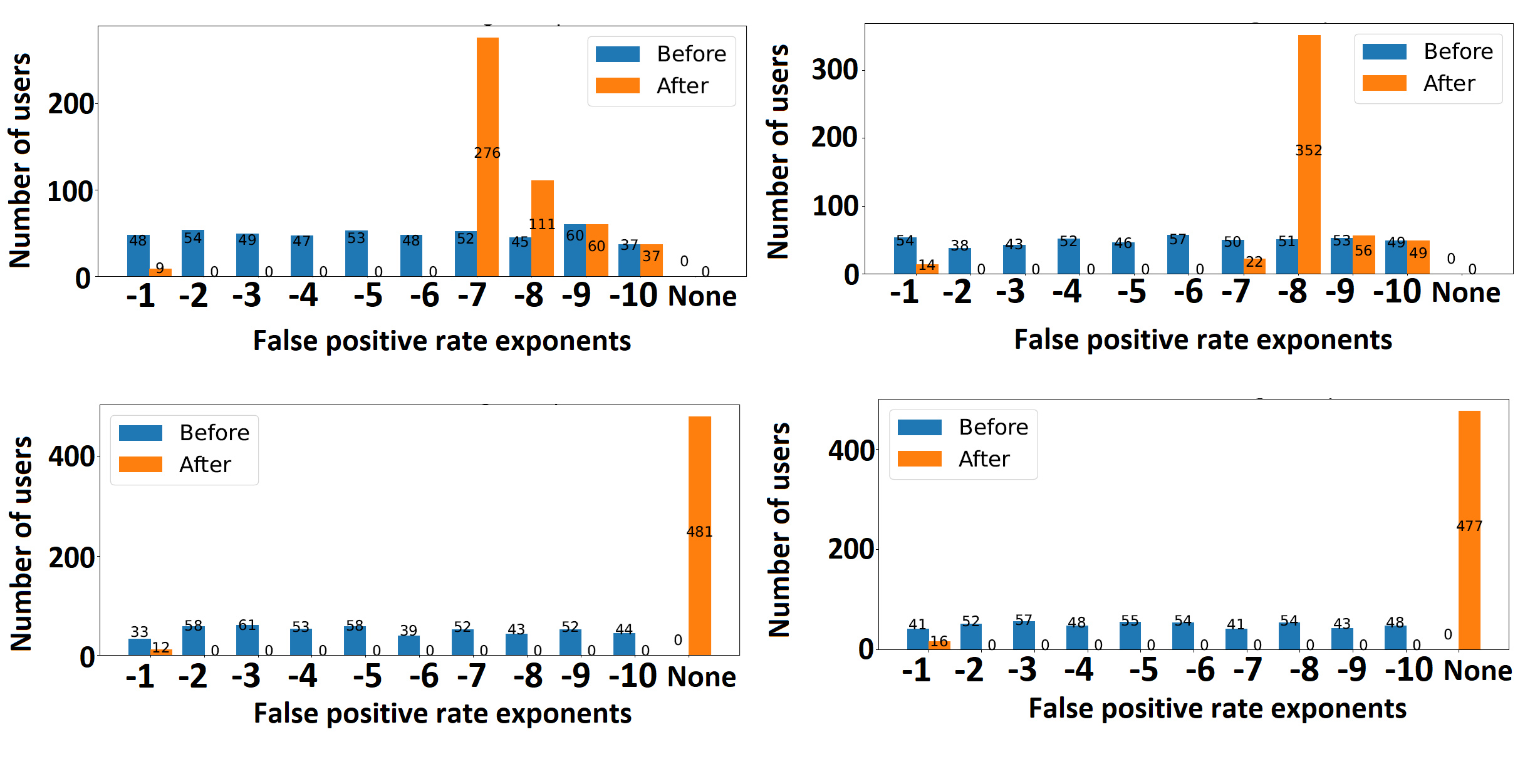}
    \caption{Strategy profile in $\varepsilon$-NE, random init., \textit{mail} dataset, top: global altruism, bottom: local a., left: $a_u=0.1$, right $a_u=1.0$.}
    \label{fig:mail_NE}
    \vspace{0.2cm}
    \includegraphics[width=0.85\textwidth]{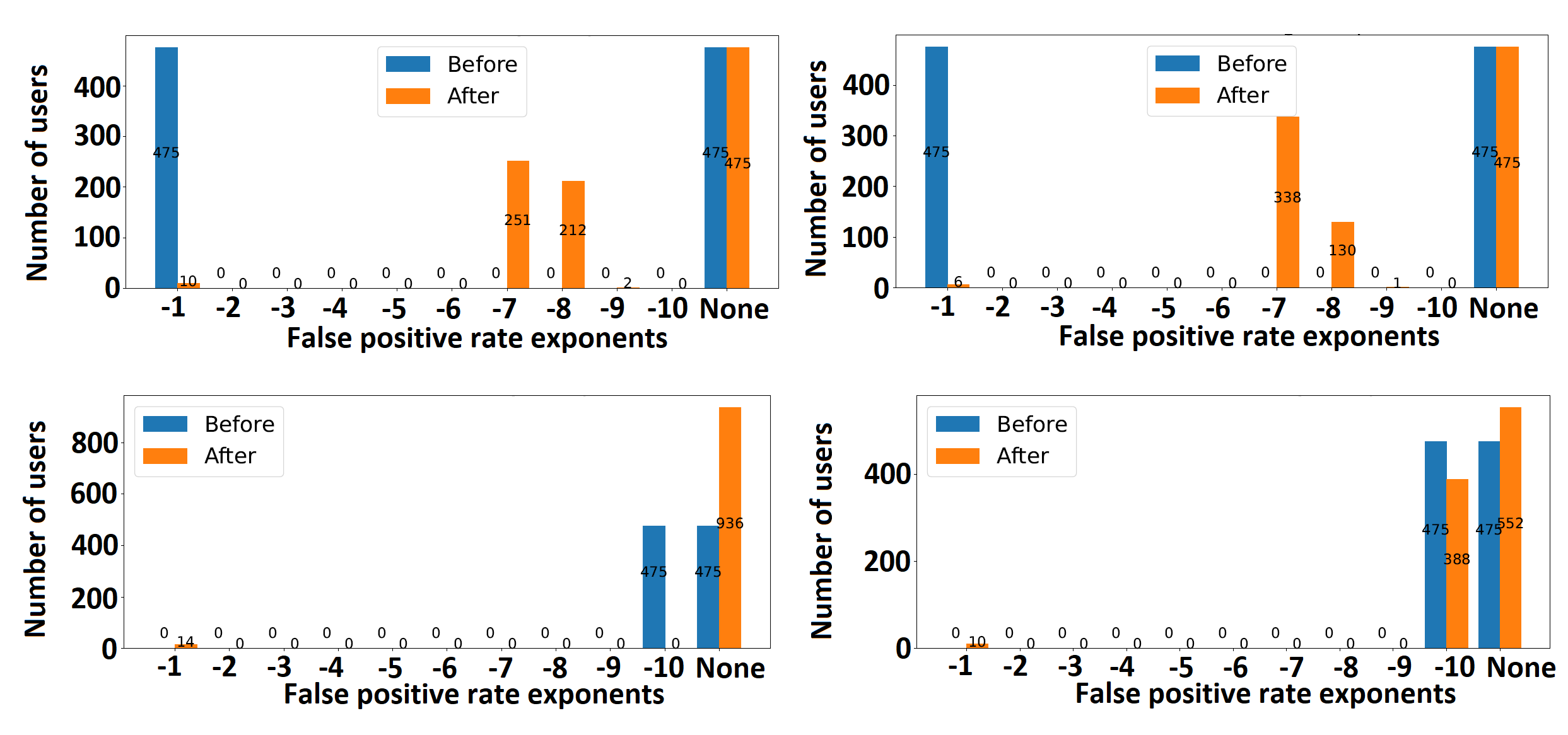}
    \caption{Strategy profile in $\varepsilon$-NE, threshold init. with node degree of $4$, \textit{message} dataset, top: init. $p_0=2^{-1}$, bottom: init. $p_0=2^{-10}$, left: $a_u=0.1$, right$a_u=1.0$.}
    \label{fig:NE_message_all}
\end{figure}
\clearpage}

\afterpage{
\begin{figure}[tb]
    \centering
    \includegraphics[width=0.99\textwidth]{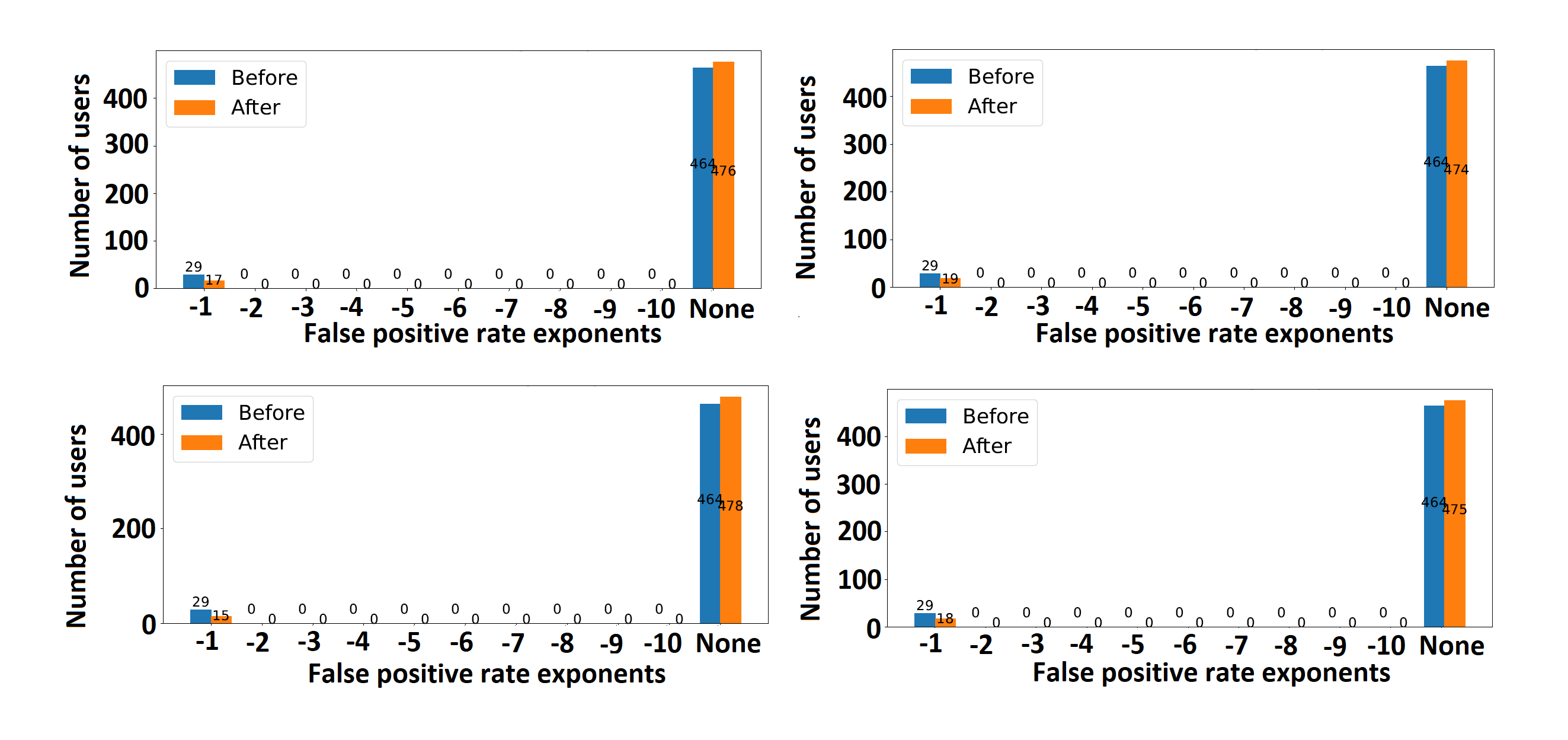}
    \caption{Strategy profile in $\varepsilon$-NE,  threshold init. with betw. centr. of $0.01$ and init. $p_0=2^{-1}$ , \textit{mail} dataset, top: global altruism, bottom: local a., left: $a_u=0.1$, right $a_u=1.0$.}
    \label{fig:NE_mail_bc}
    \renewcommand{\subfigurename}{Fig.}
    \renewcommand{\thefigure}{}
    \begin{subfigure}[b]{0.48\textwidth}
        \centering
        \renewcommand{\thesubfigure}{8}
        \includegraphics[width=0.9\textwidth]{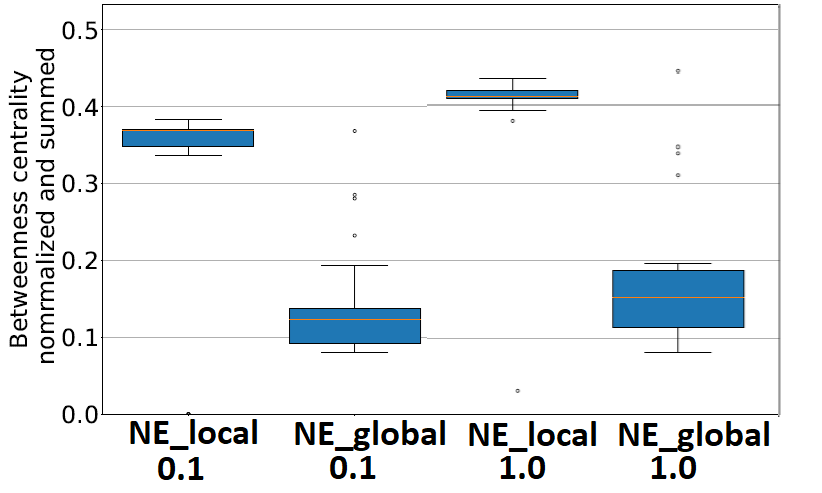}
        \caption{Aggregated betw. centr. of users with NE strategy $p_u=2^{-1}$, left: $a_u=0.1$, right:$a_u=1.0$.}
        \label{fig:bc_sum}
    \end{subfigure}
    \hfill
    \begin{subfigure}[b]{0.48\textwidth}
        \centering
        \renewcommand{\thesubfigure}{9}
        \includegraphics[width=0.9\textwidth]{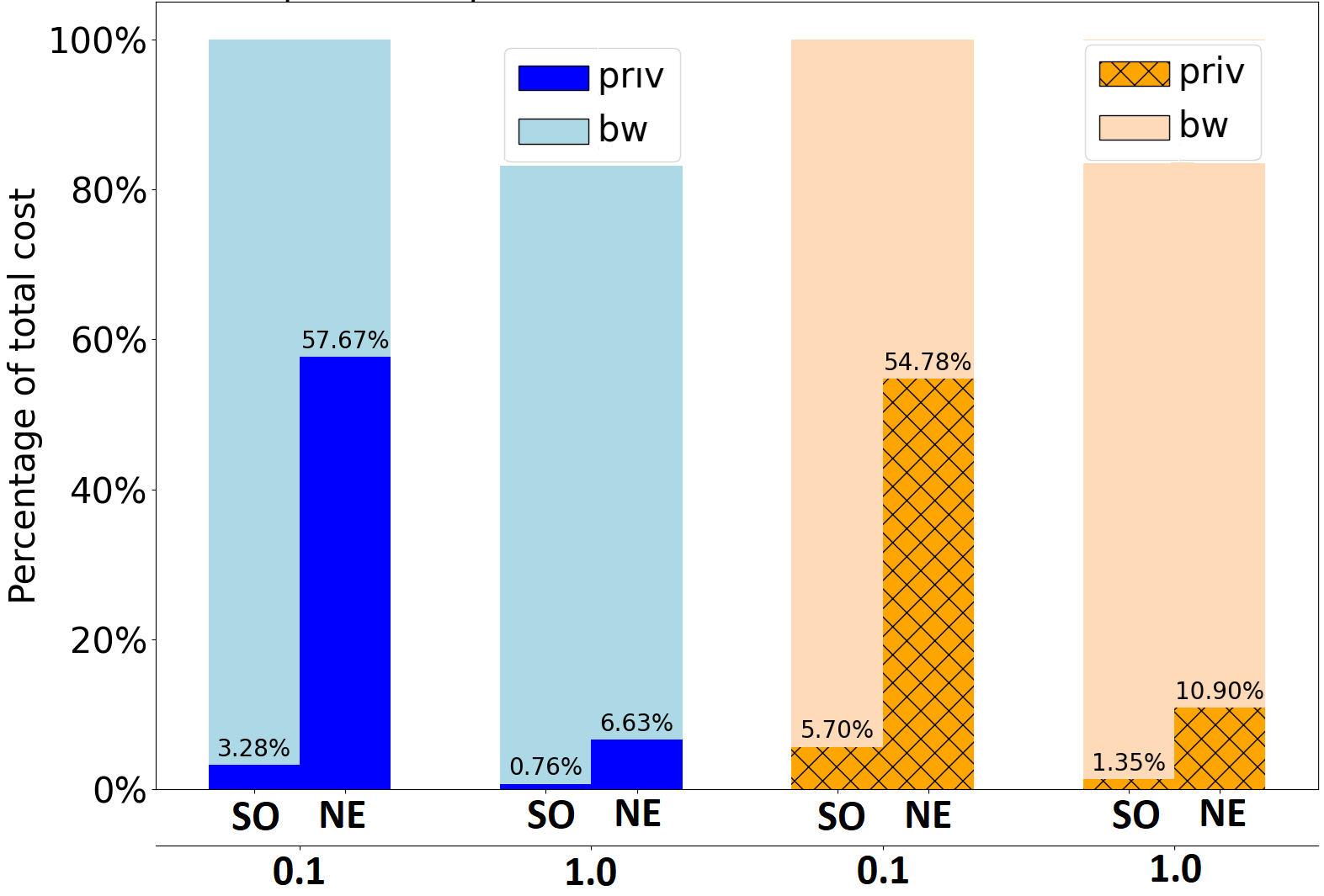}
        \caption{Composition of overall cost: privacy vs. bandwidth, global altruism, left (blue): \textit{mail}, right (orange): \textit{message}.}
        \label{fig:cost_comp_glob}
    \end{subfigure}
    \vspace{0.2cm}
    
    \renewcommand{\thefigure}{10}
    \includegraphics[width=0.99\textwidth]{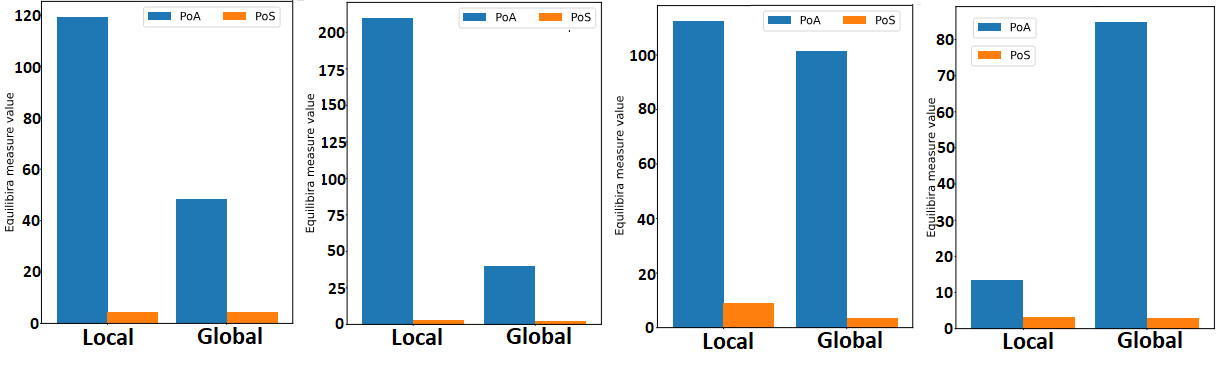}    
    \caption{Equilibrium efficiency: PoA and PoS, local and global altruism, left: \textit{mail}, right: \textit{message}, odd: $a_u=0.1$, even: $a_u=1.0$.}
    \label{fig:po}
\end{figure}
\clearpage}

\vspace{0.1cm}\noindent\textbf{Thresholding initialization (node degree). }
Recall that we expected this initialization strategy to reveal extreme equilibria. Fig.~\ref{fig:NE_message_all} shows the results for the \textit{message} dataset with global altruism using the node degree of 4 as the separation threshold. The top row corresponds to a high $p_u=2^{-1}$ and the bottom to a low $p_u=2^{-10}$ initial setting for non-selfish users, with the left column $a_u=0.1$ and the right column $a_u=1.0$.
Again, there is a small set of users in NE providing maximum cover traffic (leftmost column in every subfigure), while there are a lot of free-riders (rightmost column). When the BRD is initialized with minimal cover traffic, the NE is extremely polarized (bottom row); while with high initial settings, around half of the population is settled at medium equilibrium values. This result highlights the existence of very different equilibria in the same scenario.
Results for the other dataset and local altruism are similar in nature (not shown).

\vspace{0.1cm}\noindent\textbf{Thresholding initialization (betweeness centrality). }
Fig.~\ref{fig:NE_mail_bc} displays the results for the \textit{mail} dataset with global altruism using betweenness centrality of $0.01$ as the separation threshold. The top row corresponds to global and the bottom to local altruism, with the left column $a_u=0.1$ and the right column $a_u=1.0$. From the figure, it is evident that only a few nodes are above the threshold, and even fewer will provide maximum cover traffic in equilibrium. Note that the convergence is the shortest with this initialization method: the BRD algorithm finishes after relatively few steps. Note that the resulting equilibrium is again a polarized one, with a few players shouldering the burden of cover traffic. Also, note that we experienced similar behavior when using the same initialization strategy in other experiments.

Motivated by this result, we wanted to characterize how important the small set of nodes \emph{providing maximum equilibrium cover traffic, $p_u=2^{-1}$} is (we refer to these nodes as \emph{max nodes}). As betweenness centrality is a valid node importance measure when it comes to a communication network~\cite{bc_networks}, we computed the aggregated betweenness centrality of max nodes across all discovered equilibria for the same dataset. Fig.~\ref{fig:bc_sum} shows how these partial aggregates stack up against the network aggregate under both local and global altruism regimes (\textit{mail} dataset, normalized betweenness centrality values, left: $a_u=0.1$, right $a_u=1.0$). We can make three observations: i) max nodes (sometimes only $4\%$ of the population) correspond to a large proportion (up to $44\%$) of betweenness centrality, ii) local altruism results in a stronger concentration of betweenness centrality in max nodes, and iii) stronger altruism (higher $a_u$) also implies more profound concentration).

We refer the interested reader to Appendix D in our technical report~\cite{tech} for more results. Interestingly, a large majority of efficient equilibria (i.e., with low social cost) are characterized by such a strong concentration of betweenness centrality. Note that the \emph{Threshold} initialization strategy based on betweenness centrality resulted in equilibria with low social cost and the shortest convergence time.

\subsection{Equilibrium versus Social Optimum}

\begin{table}[!t]
    \centering
    \setlength{\arrayrulewidth}{0.5mm}
    \setlength{\tabcolsep}{8pt} 
    \renewcommand{\arraystretch}{1.2} 
    \begin{adjustbox}{max width=\textwidth}
    \begin{tabular}{|c|c|c|c|c|c|c|c|c|c|}
        \hline
        \textbf{DS} & \textbf{A} & \textbf{Solution} & \textbf{A. model} & \textbf{Priv Cost} & \textbf{BW Cost} & \textbf{BC 10th\%} & \textbf{BC 50th\%} & \textbf{BC 90th\%} & \textbf{Initial setup} \\
        \hline
        \multirow{8}{*}{\rotatebox[origin=c]{90}{Message}} & \multirow{4}{*}{0.1} & \multirow{2}{*}{SO} & Local & -7694.99 & -127389.5 & 0.8343 & 1.1985 & 1.2108 & bc, Threshold, \newline all from -10 \\
        & & & Global & -3880.18 & -134877.0 & 0.8343 & 1.1985 & 1.2108 & bc, Threshold, \newline all from -10 \\
        \cline{3-10}
        & & \multirow{2}{*}{NE} & Local & -427417.87 & -82507.0 & 0.8343 & 1.1985 & 1.2108 & random 2 \\
        & & & Global & -117871.68 & -97305.0 & 0.8343 & 1.1985 & 1.2108 & bc, Threshold, \newline all from -10  \\
        \cline{2-10}
        & \multirow{4}{*}{1.0} & \multirow{2}{*}{SO} & Local & -1948.76 & -142394.5 & 0.8343 & 1.1985 & 1.2108 & 'bc', 'Threshold', \newline 'all from -10' \\
        & & & Global & -501.74 & -157243.0 & 0.8343 & 1.1985 & 1.2108 & 'bc', 'Threshold', \newline 'all from -1' \\
        \cline{3-10}
        & & \multirow{2}{*}{NE} & Local & -29810.20 & -112517.0 & 0.8343 & 1.1985 & 1.2108 & random 0 \\
        & & & Global & -14699.95 & -120205.5 & 0.8085 & 1.1985 & 1.2108 & 'degree', 'Threshold', \newline'all from -10' \\
        \hline
        \multirow{8}{*}{\rotatebox[origin=c]{90}{Mail}} & \multirow{4}{*}{0.1} & \multirow{2}{*}{SO} & Local & -26872.01 & -792918.5 & 0.6891 & 1.1407 & 1.2064 & 'bc', 'Threshold', \newline 'all from -10' \\
        & & & Global & -13757.53 & -831782.0 & 0.6329 & 1.1407 & 1.2064 & 'bc', 'Threshold',\newline 'all from -1' \\
        \cline{3-10}
        & & \multirow{2}{*}{NE} & Local & -1540593.96 & -554842.5 & 0.6508 & 1.1407 & 1.2064 & random 5 \\
        & & & Global & -809226.32 & -594074.0 & 0.65 & 1.1407 & 1.2064 & 'bc', 'Threshold', \newline'all from -10' \\
        \cline{2-10}
        & \multirow{4}{*}{1.0} & \multirow{2}{*}{SO} & Local & -6718.36 & -873060.3 & 0.7233 & 1.1407 & 1.2064 & 'bc', 'Threshold',\newline 'all from -10' \\
        & & & Global & -1740.72 & -950986.5 & 0.6802 & 1.1407 & 1.2064 & 'bc', 'Threshold', \newline'all from -1' \\ 
        \cline{3-10}
        & & \multirow{2}{*}{NE} & Local & -107338.36 & -712944.0 & 0.6809 & 1.1407 & 1.2064 & random 8\\
        & & & Global & -53449.33 & -753186.0 & 0.6644 & 1.1407 & 1.2064 & 'bc', 'Threshold', \newline'all from -10'\\
        \hline
    \end{tabular}
    \end{adjustbox}
    \caption{``Empirical CDF'' percentiles of aggregated betw. centr. and costs over users ordered by decreasing cover traffic for SO and best-case NE.}
    \label{tab:best}
\end{table}

\begin{figure}[tb]
    \centering
    \includegraphics[width=0.6\textwidth]{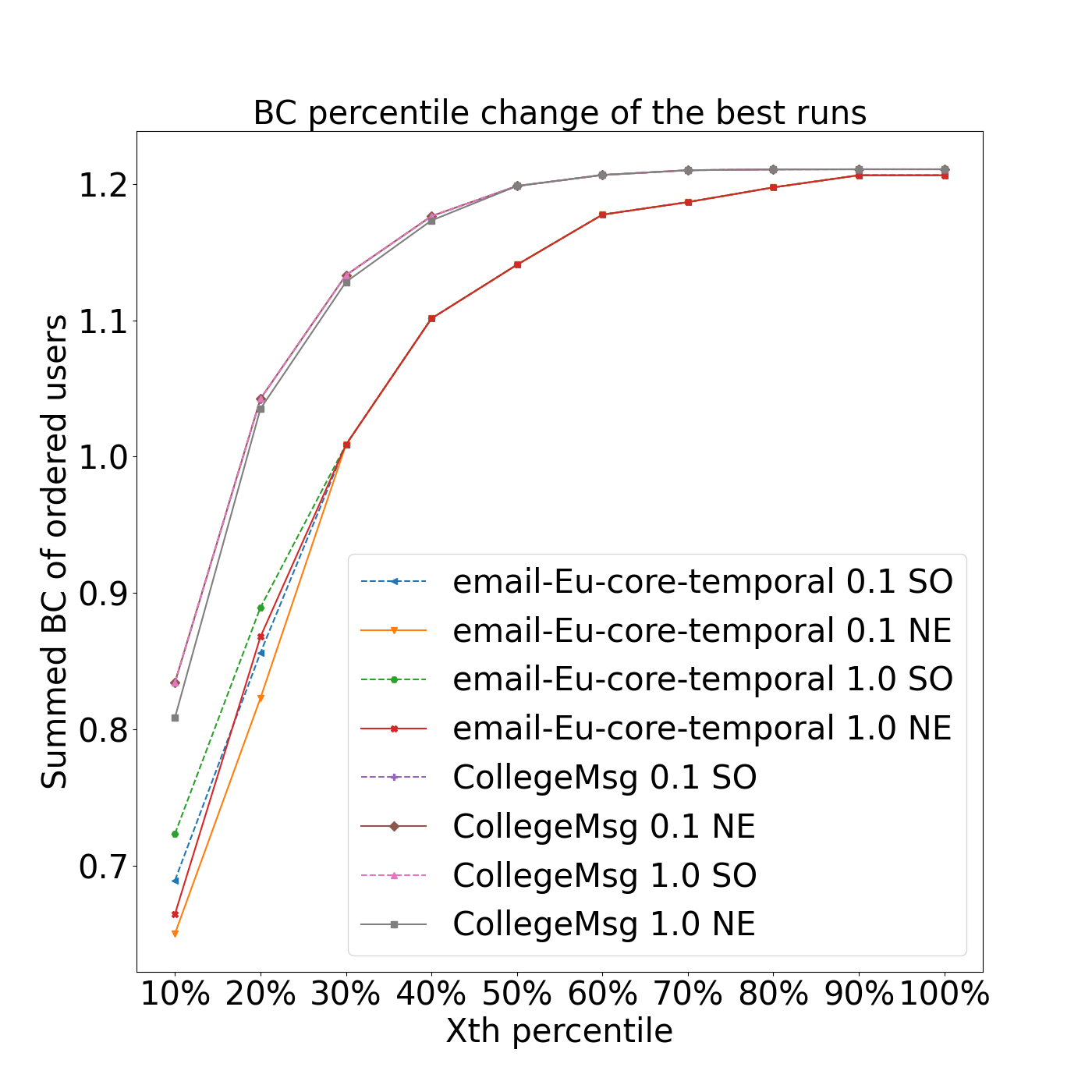}
    \caption{``Empirical CDF'' of aggregated betw. centr. over users ordered by decreasing cover traffic for SO and best-case NE.}
    \label{fig:best_runs_plot_BC}
\end{figure}

The defining difference between the equilibrium and social optimum strategy profiles is the presence (or absence) of free-riders. Also, SO corresponds to a larger amount of aggregate cover traffic.

Another somewhat expected result is the change in the composition of overall cost: while in an NE, the privacy cost dominates (e.g., $>50\%$ in case of low altruism), at the SO, the bandwidth cost dominates ($<90\%$). This is visualized in Fig.~\ref{fig:cost_comp_glob} for the global altruism regime; left (blue) and right (orange) denote results for the \textit{mail} and \textit{message} dataset, respectively. It is also clear that stronger altruism, i.e., $a_u=1.0$, reduces the proportion of the privacy cost. In the case of local altruism, these trends are even more pronounced (see Appendix D in our technical report~\cite{tech}. 

Fig.~\ref{fig:best_runs_plot_BC} combined with Table~\ref{tab:best} show the change of concentration of aggregated betweenness centrality in the best NE and at SO over the nodes ordered by their chosen strategies (decreasing order of false positive detection rate/cover traffic). It is clear that already the top $10\%$ of contributors correspond to $60-70\%$ of total betweenness centrality, making it a decisive factor in practical optimal mechanism design. Note that the curves are similar regardless of the dataset, altruism level, or chosen BRD initialization strategy.

\vspace{0.1cm}\noindent\textbf{Price of Anarchy and  Stability. }
Price of Stability (PoS) and Price of Anarchy (PoA) are valuable for comparing NE and SO as they illustrate the efficiency of strategic decisions: PoS measures the best-case efficiency loss, while PoA assesses the worst-case loss when players act in their self-interest.
Fig.~\ref{fig:po} shows PoA and PoS values for both altruism models, where the worst and best NE were chosen from the equilibrium outcomes of all related experiments.
The leftmost (rightmost) two plots correspond to the \emph{mail} (\emph{message}) dataset, where odd (even) plots belong to low (high) $a_u$ values.

Generally speaking, PoA values are quite high, while PoS values are comparably much lower. The impact of different communication networks is also apparent: the PoA is much lower with high but local altruism in the \emph{message} network compared to the \emph{mail} network (fourth plot vs. second plot), while the opposite is true for high global altruism. We leave it to future work, whether this can be attributed to the density difference between the two graphs.

These results imply that in order for the FMD system to achieve near-optimal or optimal operation, a mechanism designer (e.g., the app developer) should either i) be able to set the cover traffic parameters centrally or ii) re-design the system incentives in a way to elicit a favorable equilibrium, similar to the best-case NE in our experiments. With regard to the central optimal design, one should perform a network analysis to determine the participants' betweenness centrality parameter and set the cover traffic parameters accordingly (see Fig.~ \ref{fig:college_SO}).



%% file: 6conclusion.tex
\section{Conclusion}
\label{sec:conc}

Fuzzy Message Detection has attracted significant academic and some commercial interest since its inception. However, Seres et. al.~\cite{seres2022effect} raised some concerns about the privacy guarantees that the Fuzzy Message Detection~\cite{beck2021fuzzy} scheme, gaining popularity and is being integrated into real-world apps, provides. They conjectured that with selfish users, the system is not viable. In contrast, in this paper, we showed that the presence of a few altruistic users may alleviate this situation and yield a viable equilibrium. 
By means of empirical game-theoretical analysis, utilizing real-world communication datasets, we i) characterized the emerging equilibria, ii) quantified the impact of different types and levels of altruism, and iii) assessed the efficiency of potential outcomes versus socially optimal allocations. Furthermore, taking a mechanism design approach, we showed how the betweenness centrality measure could be utilized to achieve the social optimum.



\noindent\textbf{Practical considerations. }
It is not trivial how a messaging app provider can facilitate (near-)optimal operation in a real-life deployment. First of all, the system is dynamic, with nodes and communication patterns changing, even on short timescales. Second, directly computing betweenness centrality is not feasible owing to the inherent properties of the anonymous messaging technology. However, a solution based on secure multi-party computation~\cite{smc_bc} could be integrated into the messaging logic. Third, the provider could set favorable homogeneous default values (see Figure \ref{fig:swcentral}), which then would be left unchanged with high probability~\cite{defaults_priv}.

\noindent\textbf{Limitations and Future Work. }We have barely scratched the surface of altruistic anonymous messaging systems. First, we investigated a simple model where players have perfect knowledge and perfect rationality and play a one-shot game. We only provided empirical results, limited the parameter space, and restricted ourselves to two datasets in order to cope with computational constraints. Bounded rationality, imperfect knowledge, and the temporal dynamics of users and communication patterns could call for more sophisticated modeling and simulation studies. Furthermore, concepts from cooperative game theory could be used to integrate rewards into the system.

Second, we decided to focus our analysis on FMD (and its claimed relationship anonymity), a promising technology gaining popularity. While the utility functions studied are FMD-specific, we believe our approach could be generalized to other anonymous messaging/payment protocols, also providing group communication functionality. Even more ambitious and potentially more impactful, we may be able to generalize our empirical game-theoretical analysis to any hiding-in-the-crowd type privacy-preserving mechanism where a user's privacy inherently depends on other users' actions~\cite{biczok2013interdependent,humbert2019survey}.


%% file: main.bbl
\begin{thebibliography}{10}
\providecommand{\url}[1]{\texttt{#1}}
\providecommand{\urlprefix}{URL }
\providecommand{\doi}[1]{https://doi.org/#1}

\bibitem{freeriding_gnutella}
Adar, E., Huberman, B.A.: Free riding on gnutella. First monday  (2000)

\bibitem{article:priceofstability}
Anshelevich, E., Dasgupta, A., Kleinberg, J., Tardos, {\'E}., Wexler, T.,
  Roughgarden, T.: The price of stability for network design with fair cost
  allocation. SIAM Journal on Computing  \textbf{38}(4),  1602--1623 (2008)

\bibitem{article:selfishness-level}
Apt, K., Schaefer, G.: Selfishness level of strategic games. Annals of Pure and
  Applied Logic - APAL  \textbf{49} (05 2011)

\bibitem{altruism1}
Barclay, P.: Trustworthiness and competitive altruism can also solve the
  “tragedy of the commons”. Evolution and Human Behavior  \textbf{25}(4),
  209--220 (2004). \doi{https://doi.org/10.1016/j.evolhumbehav.2004.04.002},
  \url{https://www.sciencedirect.com/science/article/pii/S109051380400025X}

\bibitem{beck2021fuzzy}
Beck, G., Len, J., Miers, I., Green, M.: Fuzzy message detection. In:
  Proceedings of the 2021 ACM SIGSAC Conference on Computer and Communications
  Security. pp. 1507--1528 (2021)

\bibitem{biczok2013interdependent}
Bicz{\'o}k, G., Chia, P.H.: Interdependent privacy: Let me share your data. In:
  International conference on financial cryptography and data security. pp.
  338--353. Springer (2013)

\bibitem{chor1998private}
Chor, B., Kushilevitz, E., Goldreich, O., Sudan, M.: Private information
  retrieval. Journal of the ACM (JACM)  \textbf{45}(6),  965--981 (1998)

\bibitem{bc_networks}
Dolev, S., Elovici, Y., Puzis, R.: Routing betweenness centrality. J. ACM
  \textbf{57}(4) (may 2010). \doi{10.1145/1734213.1734219},
  \url{https://doi.org/10.1145/1734213.1734219}

\bibitem{feldman2006free}
Feldman, M., Papadimitriou, C., Chuang, J., Stoica, I.: Free-riding and
  whitewashing in peer-to-peer systems. IEEE Journal on Selected Areas in
  Communications  \textbf{24}(5),  1010--1019 (2006)

\bibitem{fletcher2007evolution}
Fletcher, J.A., Zwick, M.: The evolution of altruism: Game theory in multilevel
  selection and inclusive fitness. Journal of theoretical biology
  \textbf{245}(1),  26--36 (2007)

\bibitem{tech}
Frank, M., Pejo, B., Biczok, G.: {Effective Anonymous Messaging: the Role of
  Altruism (technical report)} (2024), \url{https://cloud.crysys.hu/s/fmdgt}

\bibitem{freeman1977set}
Freeman, L.C.: A set of measures of centrality based on betweenness. Sociometry
  pp. 35--41 (1977)

\bibitem{freeriding_napster}
Golle, P., Leyton-Brown, K., Mironov, I.: Incentives for sharing in
  peer-to-peer networks. In: Proceedings of the 3rd ACM conference on
  Electronic Commerce. pp. 264--267 (2001)

\bibitem{altruism2}
Greenwood, G.W.: Altruistic punishment can help resolve tragedy of the commons
  social dilemmas. In: 2016 IEEE Conference on Computational Intelligence and
  Games (CIG). pp.~1--7 (2016). \doi{10.1109/CIG.2016.7860402}

\bibitem{hardin_toc}
Hardin, G.: The tragedy of the commons. Science  \textbf{162}(3859),
  1243--1248 (1968). \doi{10.1126/science.162.3859.1243},
  \url{https://www.science.org/doi/abs/10.1126/science.162.3859.1243}

\bibitem{huang2020exploratory}
Huang, J., Talbi, R., Zhao, Z., Boucchenak, S., Chen, L.Y., Roos, S.: An
  exploratory analysis on users’ contributions in federated learning. In:
  2020 Second IEEE International Conference on Trust, Privacy and Security in
  Intelligent Systems and Applications (TPS-ISA). pp. 20--29. IEEE (2020)

\bibitem{humbert2019survey}
Humbert, M., Trubert, B., Huguenin, K.: A survey on interdependent privacy. ACM
  Computing Surveys (CSUR)  \textbf{52}(6),  1--40 (2019)

\bibitem{p2p_incentives_survey2023}
Ihle, C., Trautwein, D., Schubotz, M., Meuschke, N., Gipp, B.: Incentive
  mechanisms in peer-to-peer networks — a systematic literature review. ACM
  Comput. Surv.  \textbf{55}(14s) (jul 2023). \doi{10.1145/3578581},
  \url{https://doi.org/10.1145/3578581}

\bibitem{jakkamsetti2023scalable}
Jakkamsetti, S., Liu, Z., Madathil, V.: Scalable private signaling. Cryptology
  ePrint Archive  (2023)

\bibitem{tor_incentives}
``Johnny''~Ngan, T.W., Dingledine, R., Wallach, D.S.: Building incentives into
  tor. In: Sion, R. (ed.) Financial Cryptography and Data Security. pp.
  238--256. Springer Berlin Heidelberg, Berlin, Heidelberg (2010)

\bibitem{jun2005incentives}
Jun, S., Ahamad, M.: Incentives in bittorrent induce free riding. In:
  Proceedings of the 2005 ACM SIGCOMM workshop on Economics of peer-to-peer
  systems. pp. 116--121 (2005)

\bibitem{freeriding_p2p}
Karakaya, M., Korpeoglu, I., Ulusoy, O.: Free riding in peer-to-peer networks.
  IEEE Internet Computing  \textbf{13}(2),  92--98 (2009).
  \doi{10.1109/MIC.2009.33}

\bibitem{article:priceofanarchy}
Koutsoupias, E., Papadimitriou, C.H.: Worst-case equilibria. Comput. Sci. Rev.
  \textbf{3},  65--69 (1999)

\bibitem{smc_bc}
Kukkala, V.B., Iyengar, S.R.S.: Computing betweenness centrality: An efficient
  privacy-preserving approach. In: Camenisch, J., Papadimitratos, P. (eds.)
  Cryptology and Network Security. pp. 23--42. Springer International
  Publishing, Cham (2018)

\bibitem{kunreuther2003interdependent}
Kunreuther, H., Heal, G.: Interdependent security. Journal of risk and
  uncertainty  \textbf{26},  231--249 (2003)

\bibitem{laszka_interdepsec}
Laszka, A., Felegyhazi, M., Buttyan, L.: A survey of interdependent information
  security games. ACM Comput. Surv.  \textbf{47}(2) (aug 2014).
  \doi{10.1145/2635673}, \url{https://doi.org/10.1145/2635673}

\bibitem{lewis2021niwl}
Lewis, S.J.: Niwl: a prototype system for open, decentralized, metadata
  resistant communication using fuzzytags and random ejection mixers (2021),
  \url{\url{https://git.openprivacy.ca/openprivacy/niwl}}

\bibitem{liu2022oblivious}
Liu, Z., Tromer, E.: Oblivious message retrieval. In: Advances in
  Cryptology--CRYPTO 2022: 42nd Annual International Cryptology Conference,
  CRYPTO 2022, Santa Barbara, CA, USA, August 15--18, 2022, Proceedings, Part
  I. pp. 753--783. Springer (2022)

\bibitem{liu2023group}
Liu, Z., Tromer, E., Wang, Y.: Group oblivious message retrieval. Cryptology
  ePrint Archive  (2023)

\bibitem{madathil2022private}
Madathil, V., Scafuro, A., Seres, I.A., Shlomovits, O., Varlakov, D.: Private
  signaling. In: 31st USENIX Security Symposium (USENIX Security 22). pp.
  3309--3326 (2022)

\bibitem{article:Shapley1994PotentialG}
Monderer, D., Shapley, L.S.: Potential games. Games and economic behavior
  \textbf{14}(1),  124--143 (1996)

\bibitem{nash_gt_original}
Nash, J.: Non-cooperative games. Annals of mathematics pp. 286--295 (1951)

\bibitem{collegemsg}
Panzarasa, P., Opsahl, T., Carley, K.: Patterns and dynamics of users' behavior
  and interaction: Network analysis of an online community. JASIST
  \textbf{60},  911--932 (05 2009). \doi{10.1002/asi.21015}

\bibitem{eumessage}
Paranjape, A., Benson, A.R., Leskovec, J.: Motifs in temporal networks. In:
  Proceedings of the Tenth ACM International Conference on Web Search and Data
  Mining. p. 601–610. WSDM '17, Association for Computing Machinery, New
  York, NY, USA (2017). \doi{10.1145/3018661.3018731},
  \url{https://doi.org/10.1145/3018661.3018731}

\bibitem{pejo_covid}
Pej\'{o}, B., Bicz\'{o}k, G.: Games in the time of covid-19: Promoting
  mechanism design for pandemic response. ACM Trans. Spatial Algorithms Syst.
  \textbf{8}(3) (jan 2022). \doi{10.1145/3503155},
  \url{https://doi.org/10.1145/3503155}

\bibitem{rondelet2021zeth}
Rondelet, A.: Fuzzy message detection in zeth (2021),
  \url{\url{https://github.com/clearmatics/zeth-specifications/issues/18}}

\bibitem{roughgarden_2016}
Roughgarden, T.: Twenty Lectures on Algorithmic Game Theory. Cambridge
  University Press (2016). \doi{10.1017/CBO9781316779309}

\bibitem{mental_idp}
Schoenherr, J.: Whose privacy, what surveillance? dimensions of the mental
  models for privacy and security. IEEE Technology and Society Magazine
  \textbf{41}(1),  54--65 (2022). \doi{10.1109/MTS.2022.3147536}

\bibitem{seres2022effect}
Seres, I.A., Pej{\'o}, B., Burcsi, P.: The effect of false positives: Why fuzzy
  message detection leads to fuzzy privacy guarantees? In: International
  Conference on Financial Cryptography and Data Security. pp. 123--148.
  Springer (2022)

\bibitem{simon1993altruism}
Simon, H.A.: Altruism and economics. The American Economic Review
  \textbf{83}(2),  156--161 (1993)

\bibitem{devalence2021penumbra}
de~Valence, H.: Determine whether penumbra could integrate fuzzy message
  detection (2021),
  \url{\url{https://github.com/penumbra-zone/penumbra/issues/4}}

\bibitem{varian2004system}
Varian, H.: System reliability and free riding. In: Economics of information
  security, pp. 1--15. Springer (2004)

\bibitem{defaults_priv}
Watson, J., Lipford, H.R., Besmer, A.: Mapping user preference to privacy
  default settings. ACM Trans. Comput.-Hum. Interact.  \textbf{22}(6) (nov
  2015). \doi{10.1145/2811257}, \url{https://doi.org/10.1145/2811257}

\bibitem{DBLP:conf/aaai/Wellman06}
Wellman, M.P.: Methods for empirical game-theoretic analysis. In: Proceedings,
  The Twenty-First National Conference on Artificial Intelligence and the
  Eighteenth Innovative Applications of Artificial Intelligence Conference,
  July 16-20, 2006, Boston, Massachusetts, {USA}. pp. 1552--1556. {AAAI} Press
  (2006), \url{http://www.aaai.org/Library/AAAI/2006/aaai06-248.php}

\bibitem{DBLP:conf/ijcai/Yu0V21}
Yu, S., Kempe, D., Vorobeychik, Y.: Altruism design in networked public goods
  games. In: Zhou, Z. (ed.) Proceedings of the Thirtieth International Joint
  Conference on Artificial Intelligence, {IJCAI} 2021, Virtual Event /
  Montreal, Canada, 19-27 August 2021. pp. 493--499. ijcai.org (2021).
  \doi{10.24963/IJCAI.2021/69}, \url{https://doi.org/10.24963/ijcai.2021/69}

\bibitem{gpath_tor}
Zhang, N., Yu, W., Fu, X., Das, S.K.: gpath: A game-theoretic path selection
  algorithm to protect tor’s anonymity. In: International Conference on
  Decision and Game Theory for Security. pp. 58--71. Springer (2010)

\end{thebibliography}
